\documentclass{article}

\usepackage{PRIMEarxiv}

\usepackage[utf8]{inputenc} 
\usepackage[T1]{fontenc}    
\usepackage{hyperref}       
\usepackage{url}            
\usepackage{booktabs}       
\usepackage{amsfonts}       
\usepackage{nicefrac}       
\usepackage{microtype}      
\usepackage{lipsum}
\usepackage{fancyhdr}       
\usepackage{graphicx}       
\graphicspath{{media/}}     
\usepackage{xcolor}
\usepackage{setspace}
\usepackage[most]{tcolorbox} 
\usepackage{enumitem}
\usepackage{caption} 
\usepackage{wrapfig} 
\usepackage[numbers]{natbib}
\setcitestyle{square,comma,numbers,maxnames=3}

\title{Practitioners' Perspectives on a Differential Privacy Deployment Registry
}

\author{
  Priyanka Nanayakkara, Elena Ghazi, Salil Vadhan \\
  OpenDP, Harvard University \\
  \texttt{\{priyankan, elenaghazi, svadhan\}@g.harvard.edu} \\
}

\begin{document}

\maketitle

\vspace{-7mm}
\begin{center}
    \today{}
\end{center}
\vspace{5mm}

\begin{abstract}
Differential privacy (DP)---a principled approach to producing statistical data products (e.g., summary statistics, machine learning models) with strong, mathematically provable privacy guarantees for the individuals in the underlying dataset---has seen substantial adoption in practice over the past decade. Applying DP requires making several implementation decisions, each with significant impacts on data privacy and/or utility. Hence, to promote shared learning and accountability around DP deployments, Dwork, Kohli, and Mulligan \cite{dwork2019differential} proposed a public-facing repository (``registry'') of DP deployments. The DP community has recently started to work toward realizing this vision. We contribute to this effort by (1) developing a holistic, hierarchical schema to describe any given DP deployment and (2) designing and implementing an interactive interface to act as a registry where practitioners can access information about past DP deployments. We (3) populate our interface with 21 real-world DP deployments and (4) conduct an exploratory user study with DP practitioners ($n=16$) to understand how they would use the registry, as well as what challenges and opportunities they foresee around its adoption. We find that participants were enthusiastic about the registry as a valuable resource for evaluating prior deployments and making future deployments. They also identified several opportunities for the registry, including that it can become a ``hub'' for the community and support broader communication around DP (e.g., to legal teams). At the same time, they identified challenges around the registry gaining adoption, including the effort and risk involved with making implementation choices public and moderating the quality of entries. Based on our findings, we offer recommendations for encouraging adoption and increasing the registry's value not only to DP practitioners, but also to policymakers, data users, and data subjects.
\end{abstract}


\section{Introduction}
Differential privacy (DP)~\cite{dwork2006calibrating} was first introduced in 2006 as a theoretical standard of privacy for computing statistical releases from datasets with sensitive information about individuals. In the nearly two decades since, researchers and practitioners have undertaken substantial work to bring DP to practical fruition. To date, DP has been deployed to produce data products ranging from summary statistics to machine learning models, by organizations such as the U.S. Census Bureau~\cite{abowd2018us}, Google~\cite{aktay2020google}, Apple~\cite{apple2017}, and the Wikimedia Foundation~\cite{adeleye2023publishing}. A recent survey found at least 42 deployments reported in papers or blog posts~\cite{khavkin2025differential}.

Deploying DP involves making a range of implementation choices, each with potential to significantly impact data privacy and/or utility. Under DP, statistical noise is injected into analyses to obscure individual-level contributions while preserving overall signal in the data. The ``privacy loss parameter'' $\epsilon$ often receives much attention for calibrating how much noise is added---and therefore controlling the privacy-accuracy tradeoff---but it is important to note that there are several other decisions (e.g., whether the deployment model is central, local, etc.; the choice of privacy unit; etc.) that are consequential to resulting guarantees, as captured by a recent set of guidelines on evaluating DP guarantees~\cite{NIST.SP.800-226}.

The public---including researchers and practitioners---is informed about real-world DP deployments, and their implementation choices, primarily through technical papers, blog posts, and talks published by data curators (i.e., organizations responsible for the deployment). These artifacts describe deployments in their own unique ways, following varying conventions and including varying degrees of detail. Furthermore, they are not disseminated in a consistent way; instead, they are shared through a mix of peer-reviewed research papers, blog posts on organizations' websites, talks at conferences, etc. In short, critical information about DP deployments is currently \textit{decentralized} and \textit{non-standardized}.

As a result, it is difficult for practitioners to find details about past deployments. The lack of shared information is especially problematic given the relative nascency of DP as a practical tool: not having a resource for shared information is a significant barrier to the development of standards around implementation choices that have critical impacts on both data privacy and utility. Moreover, the community lacks consensus around how to document a deployment, further limiting the extent to which potential standards can emerge.

To promote shared learning and overcome these challenges, Dwork, Kohli, and Mulligan~\cite{dwork2019differential} proposed the idea of a public-facing repository of DP deployments. They argued that it should contain both technical implementation decisions and sociotechnical factors, such as rationale for said decisions and the processes by which they were made. Appetite for a registry has grown in the privacy community, but the original vision has not yet been achieved. Furthermore, if a registry were to exist, it would raise several challenges around adoption and governance, which are difficult to assess without a working registry.

Thus, the community has started to take steps toward building a registry and realizing the vision of Dwork, Kohli, and Mulligan~\cite{dwork2019differential}. First, in 2021, Desfontaines~\cite{desfontainesblog20211001} created a blog post that lists several real-world deployments of DP and is periodically updated by Desfontaines with new deployments. Though the post has become a fairly well-known, trusted resource in the DP community, its focus is primarily on summarizing deployments at a high level and listing privacy-loss parameters, omitting several factors that are crucial for fully understanding a DP deployment. Furthermore, the blog post medium does not scale to larger numbers of deployments and relies on a single maintainer rather than being community sourced.

More recently, in spring 2024, a team at the privacy start-up Oblivious (Berrios, Usynin, Fitzsimons~\cite{original_registry}) started to work toward a more systematic, community-driven registry to be further developed in collaboration with OpenDP, the open-source software project of which we are a part. They created an initial prototype that was also in a blog post format, but with a table of deployments and educational content about DP. The table captured a basic set of attributes of a DP deployment (privacy-loss parameters, privacy unit, deployment model, etc.) in a non-hierarchical structure. Our work builds on and contributes to this effort.

\subsection{Our contributions}
We develop a newer, richer prototype of a DP deployment registry geared toward practitioners. Specifically:

\begin{itemize}
    \item \textbf{We propose a schema for describing DP deployments}. It builds on attributes from the registry initiated by Oblivious, and is based on existing literature that aims to synthesize DP concepts~\cite{NIST.SP.800-226, gaboardi2020programming, bailie2025topics, dwork2019differential, cowan2024hands}, our knowledge as DP researchers, and iterative feedback from DP experts. 
    
    \item \textbf{We design and implement a prototype registry as an interactive interface to display information about many deployments}, each described as per the abovementioned schema. The interface is designed to enable a user to (1) understand constituent components of a deployment, (2) find technical \& sociotechnical information about a deployment, (3) identify and compare similar deployments, and (4) systematically explore patterns across deployments.

    \item \textbf{We populate the registry with 21 real-world deployments} using the schema we developed. We base these entries off existing papers, blog posts, and other public resources describing past deployments.
\end{itemize}

Much of these contributions were incorporated back into the registry prototype started by Oblivious, which was launched in September 2025 to accompany a proposal for the U.S. National Institute of Standards and Technology (NIST) to host the registry~\cite{NIST.IR.8588.2025}. Conversely, the concurrent work on the prototype started by Oblivious has influenced our work in several ways.

Next, we conduct a user study ($n=16$) with DP practitioners, using our prototype registry as a technology probe~\cite{hutchinson2003technology} to explore how they might use the registry in practice and what challenges or opportunities they foresee around its adoption. In particular, we ask the following research questions (RQs):
\begin{itemize}
    \item \textbf{RQ1:} How does the registry challenge or reinforce practitioners’ existing beliefs about DP and prior DP deployments?
    
    \item \textbf{RQ2:} How might practitioners use the registry to
    \begin{itemize}
        \item \textbf{(a)} analyze and/or evaluate past deployments,
        \item \textbf{(b)} inform future deployments, both in terms of making deployment decisions and establishing the suitability of DP for a specific use case, and
        \item \textbf{(c)} identify, explore, and reflect on norms in the DP ecosystem?
    \end{itemize}

    \item \textbf{RQ3:} What challenges and opportunities do practitioners envision around the registry gaining adoption?
\end{itemize}

\textbf{Summary of findings.} We find that interacting with the registry did not always change participants' overall perceptions of DP, but consistently supported them in discovering low-level details and asking new questions about prior deployments. Furthermore, we find that participants performed nuanced evaluations of prior deployments, focusing not only on technical implementation choices, but also on overall societal relevance and transparency of the deployment. When tasked with making a future deployment using the registry, participants based their implementation choices off similar deployments---but determined similarity in different ways. Participants approached trends across deployments with curiosity, but also skepticism when they did not align with their prior beliefs. Finally, participants identified several opportunities the registry could open up, including becoming a ``hub'' for the community and supporting broader communication around DP (e.g., to legal teams). At the same time, they identified challenges related to the effort and risk involved with adding new deployments to the registry and moderating quality of entries.

We end with recommendations based on our findings for encouraging adoption and increasing the registry's value to parties beyond practitioners (e.g., policymakers, data users, data subjects).

\subsection{Other related work}
There have been multiple repositories focused on privacy enhancing technologies (PETs) broadly. The PETs Repository of Use Cases by the Centre for Data Ethics and Innovation~\cite{cdeiukRepositoryCases} contains case studies of real-world or proposed deployments of PETs. DP is included, but is not the focus. Furthermore, the repository focuses on describing deployments at a high level and providing links to external write-ups. Similarly, the Repository of PETs Use Cases by the U.K. government~\cite{UK_PETS_Repository} contains real-world uses of PETs and provides high-level descriptions with links. Last, the Future of Privacy Forum's Repository for PETs~\cite{fpfRepository} contains links to case studies and documents related to regulatory activity and other topics.

There have also been efforts to increase transparency around privacy practices generally. Nutrition labels~\cite{kelley2009nutrition} are an approach to documenting and communicating organizations' data-sharing practices typically with data subjects; this approach has been extended to a DP context to convey information flows permitted under different deployment models~\cite{smart2025models}. Concurrent to our work, Dibia, Lu, Bhattacharjee, Near, and Feng~\cite{dibia2025we} developed a DP privacy label for technical audiences; our schema contains the majority of information in their label and additional details.

Also concurrent to our work, Khavkin and Toch~\cite{khavkin2025differential} conducted a systematic literature review of DP deployments across academic, government, and commercial contexts. They propose a taxonomy of ``key DP configuration factors'' and study trends across deployments. Our schema includes these factors and multiple additional details.

Finally, less immediately related to our work are proposals or methods of documenting machine learning models and datasets. Model cards~\cite{mitchell2019model} were proposed as an approach to documenting machine learning models---including intended uses and performance across different groups of people---while datasheets~\cite{gebru2021datasheets} aim to document intended uses and other details of datasets. These two approaches do not focus specifically on privacy aspects of models or datasets, and do not provide a structure for how DP deployments should be reported. However, we draw inspiration from model cards' format, and create an adaptation to the DP context.
\section{Differential privacy deployment card}
To create a registry, we first develop a schema for any given DP deployment. The schema enables deployments to be represented in a standardized form and facilitates comparisons across them. We developed the schema, summarized as a ``deployment card'' (Figure~\ref{fig:deployment_card}), by triangulating between and synthesizing recommendations from multiple sources: the original proposal for a registry by Dwork, Kohli, and Mulligan~\cite{dwork2019differential}, which lists attributes of deployments that the registry should describe; NIST's guidelines for evaluating DP guarantees by Near, Darais, Lefkovitz and Howarth~\cite{NIST.SP.800-226}; an applied guidebook for DP practitioners by Cowan, Shoemate, and Pereira~\cite{cowan2024hands}; and two approaches to describing DP flavors more generally, specifically OpenDP's programming framework by Gaboardi, Hay, and Vadhan ~\cite{gaboardi2020programming} and one by Bailie~\cite{bailie2025topics}. Our schema includes a wider range of attributes than what was originally proposed in Dwork, Kohli, and Mulligan~\cite{dwork2019differential} and is hierarchical, offering additional detail and structure. We used an iterative process to develop the schema, discussing additions within our research team and seeking feedback from three external DP experts.

\subsection{Deployment card}
\label{section:deployment_card}
Here, we introduce and describe sections of the deployment card. We propose that deployment cards can be filled out according to different ``transparency tiers'' (Tier 1, 2, or 3). Each increasing tier requires more information to be disclosed about the deployment, allowing for a tradeoff between ease of registration and greater transparency. Below, we describe which sections are recommended for each tier. Throughout, we use the term ``underlying data'' to refer to the data used to produce the data product (i.e., the data being protected). Finally, we present the deployment card as it was instantiated in our prototype to accurately contextualize our user study findings. Throughout this section, we also include footnotes with possible modifications or improvements for future integration into the deployment card and registry.
\begin{wrapfigure}[40]{r}{0.5\textwidth}
\vspace{0.2cm}
\begin{tcolorbox}[
    colframe=black,     
    colback=white,   
    boxrule=0.8pt,       
    arc=4pt,            
    top=6pt,
    bottom=6pt,
    left=6pt,
    right=6pt,
    width = 0.5\textwidth,
    title={},  
    frame hidden=false
]
\centering
{\textbf{Differential privacy deployment card}} 

\vspace{0.5em}
\raggedright
\textbf{Data product}
\begin{itemize}[itemsep=-1pt, topsep=-1pt]
    \item Name
    \item Data curator
    \item Description
    \item Intended use
    \item Publication date
    \item Region
    \item Sector
\end{itemize}

\vspace{0.5em}

\textbf{Flavor}
\begin{itemize}[itemsep=-1pt, topsep=-1pt]
    \item Flavor name (e.g., pure, approximate)
    \item Data domain
    \item Unprotected quantities
\end{itemize}

\vspace{0.5em}

\textbf{Privacy loss}
\begin{itemize}[itemsep=-1pt, topsep=-1pt]
    \item Privacy unit
    \item Privacy-loss parameters
\end{itemize}

\vspace{0.5em}

\textbf{Deployment model}
\begin{itemize}[itemsep=-1pt, topsep=-1pt]
    \item Model name (e.g., central, local)
    \item Trust assumptions
    \item Release type (one release vs. many releases)
    \item Data source (dynamic vs. static)
    \item Access type (interactive vs. non-interactive)
\end{itemize}

\vspace{0.5em}

\textbf{Accounting}
\begin{itemize}[itemsep=-1pt, topsep=-1pt]
    \item Composition
    \item Post-processing
\end{itemize}

\vspace{0.5em}

\textbf{Implementation}
\begin{itemize}[itemsep=-1pt, topsep=-1pt]
    \item Pre-processing, exploratory data analysis, hyperparameter tuning
    \item Mechanisms
    \item Justification
\end{itemize}

\vspace{0.5em}

\textbf{More information \& sources}
\begin{itemize}[itemsep=-1pt, topsep=-1pt]
    \item Links to papers, blogposts, or other resources describing the deployment
    \item Link to the data product, if available
    \item Additional notes about the deployment
\end{itemize}

\end{tcolorbox}
\vspace{-0.5em}
\captionsetup{type=figure}
\captionof{figure}{Components of a DP deployment card.}
\label{fig:deployment_card}
\end{wrapfigure}

\subsubsection{Data product (basic information)}
This section provides basic information about the DP deployment. This is the only section required for Tier 1 entries.\\
\textbf{Name.} A unique, short name identifying the deployment.\\
\textbf{Data curator.} The entity publishing or otherwise producing the data product.\looseness=-1\\
\textbf{Description.} A brief description of the data product.\\
\textbf{Intended use.} How the data product is intended to be used, and by whom.\\
\textbf{Publication year.} The year the data product was published. In deployments with many releases (described below), we suggest using the year of first publication.\\
\textbf{Region.} The region in which the underlying data were collected (i.e., location in which the data subjects are based).\\
\textbf{Sector.} The sector of the data product, such as technology, healthcare, education, government, or energy.\footnote{A future schema may also include an attribute for sector of the curator.}

\subsubsection{Flavor}
This section describes the flavor of DP employed and related attributes. We suggest that Tier 2 entries specify the flavor name, while the data domain and unprotected quantities are also recommended for Tier 3 entries.\\
\textbf{Flavor name.\footnote{For precision, rename ``Flavor name'' to ``privacy measure.''}} The name of the DP flavor (e.g., pure DP~\cite{dwork2006calibrating}, approximate DP~\cite{dwork2006our}, zero-concentrated DP~\cite{bun2016concentrated}, R{\'e}nyi DP~\cite{mironov2017renyi}, Gaussian DP~\cite{dong2022gaussian}).\\
\textbf{Data domain.} Underlying datasets ``eligible for privacy protection'', i.e., ``the actual, potential, or counterfactual datasets that are to be protected'' under DP~\cite{bailie2025topics}.\\
\textbf{Unprotected quantities.} Any quantities in the data product that are unprotected by DP (e.g., statistics computed over a dataset that are published without DP noise). These quantities are sometimes referred to as ``invariants.''

\subsubsection{Privacy loss}
This section{\footnote{Instead of creating a separate section for privacy loss, an alternative is to group privacy unit with data domain (because the privacy unit is the entity whose data changes under adjacent datasets in the data domain~\cite{bailie2025topics}) and privacy-loss parameters with the privacy measure because the privacy measure defines the meaning of the privacy-loss parameters.} describes the strength of DP guarantees. Both the privacy unit and privacy-loss parameters are required for Tier 2 entries, but we suggest providing a more detailed explanation of the privacy unit for Tier 3 entries, as described below.\\ 
\textbf{Privacy unit.} The entity being protected. For example, the privacy unit might be a person or household, or a single event (``event-level'' privacy) or the set of all events associated with a particular user in one day (``user-day'' privacy). The privacy unit ``characterizes what [is being] protected''~\cite{cowan2024hands}. More formally, it is the entity whose data changes under adjacent datasets~\cite{bailie2025topics}; it is defined by the input metric and maximum input distance~\cite{gaboardi2020programming}. Tier 3 entries should also include a precise specification of what constitutes adjacent datasets.\\
\textbf{Privacy-loss parameters.} Values for parameters like $\epsilon$, $\delta$, and $\rho$. The specific parameters that apply to a given deployment will vary according to the DP flavor used. Privacy-loss parameters express the ``intensity of protection''~\cite{bailie2025topics}.

\subsubsection{Deployment model}
This section is intended to describe the deployment model as well as other characteristics of the deployment that similarly suggest specific arrangements of trust among actors. We recommend that all of the attributes below, except for trust assumptions, be included in Tier 2 entries. Tier 3 entries should include trust assumptions plus additional detailed information related to release type and access type, described below.

\textbf{Model name.} The name of the DP deployment model (e.g., central~\cite{dwork2006calibrating}, local~\cite{dwork2006calibrating,kasiviswanathan2011can}, shuffle~\cite{cheu2019distributed, bittau2017prochlo}, federated~\cite{beimel2008distributed}). Each deployment model implies different types or levels of trust in the data curator and other relevant actors. Each model also implies different points in the deployment pipeline at which DP noise is added. For instance, in the central model, a trusted data curator collects or is otherwise permitted to access raw data from data subjects. The curator is responsible for applying DP to create a privacy-protected data product that can be shared with other parties who are not permitted to see the underlying data. In the local model~\cite{dwork2006calibrating, kasiviswanathan2011can}, the curator is \textit{not} permitted to access the underlying data. Instead, DP is applied to data at the individual level, before it is sent from each data subject to the curator. The curator may then be responsible for combining data subjects’ privacy-protected data in a meaningful way. \\
\textbf{Trust assumptions.} A description of relevant actors to the deployment, trust assumptions for each actor, and rationale for these assumptions. Relevant actors include anyone who may see the data product, even partially, as well as adversaries. This item is only required for Tier 3 entries.\\
\textbf{Release type.} Whether the data product is published once (``one release'') or many times (``many releases''). For deployments that involve one release, Tier 3 entries should state plans for future uses or publications of the underlying data, if any. For deployments that involve many releases, Tier 3 entries should include a description of the refreshment timeframe (the amount of time after which the privacy loss budget resets), how privacy loss is managed over time, and whether a fixed amount of privacy loss is allowed before the underlying data are no longer queried.\footnote{These details could alternatively appear in the accounting section, as they directly relate to how privacy loss is apportioned and analyzed.}\\
\textbf{Data source.} Refers to the static or dynamic nature of the underlying data. If the underlying data are dynamic, new underlying data come in over time and new releases are periodically made (so this is always coupled with ``Release type'' being ``many releases''). This is also known as the setting of DP under \textit{continual observation}~\cite{dwork2010differential}. On the other hand, if the underlying data are static, there is a single, unchanging dataset on which one or many releases are made.\\
\textbf{Access type.} Refers to how potential users of the data product may access it; this attribute is either ``interactive'' or ``non-interactive.'' Under interactive deployments, people with permission, like data analysts, can interactively query the underlying data under DP, and receive privacy-protected query estimates. Tier 3 entries that are interactive should also describe how the privacy loss budget is apportioned to and across analysts, and whether non-collusion between analysts is assumed.\footnote{Again, these details could appear in the accounting section.} Under non-interactive deployments, people cannot interactively query the underlying data. Instead, they must interact with the published data product as is.

\subsubsection{Accounting}
This section\footnote{An alternative is to group the accounting section within an expanded implementation \textit{and} analysis section, which would include details about how privacy loss was analyzed (i.e., accounting) and verified (through proofs, stochastic testing, etc.).} describes how privacy loss is accounted for over multiple queries to the underlying data. While this idea is most precisely captured by composition, we also include post-processing. By doing so, we emphasize that post-processed transformations do not incur additional privacy loss (as per properties afforded by DP). This section is required for Tier 3 entries only.

\textbf{Composition.} Composition describes how privacy-loss parameter values accumulate across multiple queries to the underlying data under DP. Examples of composition include sequential composition and parallel composition~\cite{cowan2024hands}. Under sequential composition, the total privacy-loss parameter equals the sum of the privacy-loss parameters ($\epsilon$) per query. Under sequential composition, the total privacy-loss parameter ($\epsilon$) equals the sum of the privacy-loss parameters per query. Parallel composition says that the total privacy-loss parameter for a set of queries applied on disjoint subsets of the data is the maximum of the privacy-loss parameter values used for the individual queries.\\
\textbf{Post-processing.} Functions applied to the data product after DP is applied. These functions do not incur additional privacy loss. However, the impact of these functions may be important to know about for downstream applications that use the data product. For example, a data curator releasing a data product that includes counts of people across geographic regions may not want to publish negative privacy-protected counts, as they are seemingly nonsensical. Instead, they may set these negative counts to zero. Data users should know about these transformations to adequately account for uncertainty in their downstream analyses.

\subsubsection{Implementation}
This section provides specifics of how the DP deployment was implemented. While the theoretical guarantees can be ascertained from previous sections, there are several implementation-specific decisions that can undermine the theoretical guarantees. This section is required for Tier 3 entries only.

\textbf{Pre-processing, exploratory data analysis, hyperparameter tuning.} Prior to applying DP, data curators may query the underlying data to inform choices for the DP deployment---such as how to set the privacy-loss parameters---or otherwise pre-process the data in various ways. Steps such as data cleaning, exploratory data analysis, and hyperparameter tuning (e.g., on privacy-loss parameters) prior to DP being applied all incur privacy loss. These steps and details of whether privacy loss was accounted for, if at all, during these phases should be described here.\\
\textbf{Mechanisms.} Mechanisms, i.e., algorithms that take the sensitive, unprotected data as input and output a privacy-protected data product. We include both low-level mechanisms (e.g., Laplace or Gaussian~\cite{dwork2014algorithmic}) as well as high-level mechanisms, such as the Top Down Algorithm~\cite{abowd20222020}.
How these mechanisms were implemented (e.g., using a particular library) should be described as well. Other security-related implementation details, like measures taken to protect against known vulnerabilities, such as floating-point and timing channel attacks~\cite{jin2022we} should also be described here. Implementation code should be linked here, if available.
\\
\textbf{Justification.} Rationale for implementation decisions made by the data curator or others. This information should answer the following questions from Dwork, Kohli, and Mulligan~\cite{dwork2019differential}: ``What were the assumptions, modelling decisions, thresholds, and subjective decisions made in determining the implementation choices above? Why is the approach a thorough test of the stated assumptions? Was the process validated and verified? If so, how?''

\subsubsection{More information and sources} This section is optional, but recommended across tiers. It includes public sources for the information in the deployment card, if available. Sources might include papers or blog posts. If the data product is publicly available, a link should be included. Finally, any miscellaneous notes about the deployment that do not fit into other sections may be included here.

\section{Differential privacy deployment registry}
We instantiated the registry as an interactive interface, built in JavaScript (using libraries such as DataTables.js\footnote{\url{https://datatables.net/}} and D3.js\footnote{\url{https://d3js.org/}}), HTML, and CSS.
We designed to be usable by technically skilled practitioners (e.g., privacy engineers or technical leads) interested in deploying DP or evaluating DP deployments. While the original proposal for a registry~\cite{dwork2019differential} suggested its value for a broader range of audiences, like policymakers, we address the needs of practitioners as a first step. As such, we designed the registry keeping the following practitioner-oriented design goals (DGs) in mind. The registry should help a user:

\begin{itemize}
    \item \textbf{DG1: Understand constituent components of a deployment.} The registry should help a user understand the set of implementation decisions required for a DP deployment and potential implications of these decisions, particularly on privacy.
    \item \textbf{DG2: Find technical \& sociotechnical information about a deployment.} The registry should help a user find technical implementation decisions about a given deployment, as well as information about how these decisions were made, the intended use of the deployment, and other sociotechnical factors. Together, this information is intended to support a user in conducting their own analyses or evaluations of deployments.
    \item \textbf{DG3: Identify and compare similar deployments.} The registry should help a user identify deployments that are similar in some way (e.g., used pure DP; deployed after 2020). It should also help a user make comparisons across these deployments. For example, the registry should facilitate learning about the type of deployment model typically employed by a particular data curator.
    \item \textbf{DG4: Systematically explore patterns across deployments.} The registry should enable a user to explore trends across deployments, taking a bird's-eye view on the registry. The registry should also help the user understand trends in deployments over time.
\end{itemize}

\begin{figure}[t]
    \centering
    \includegraphics[width=\textwidth]{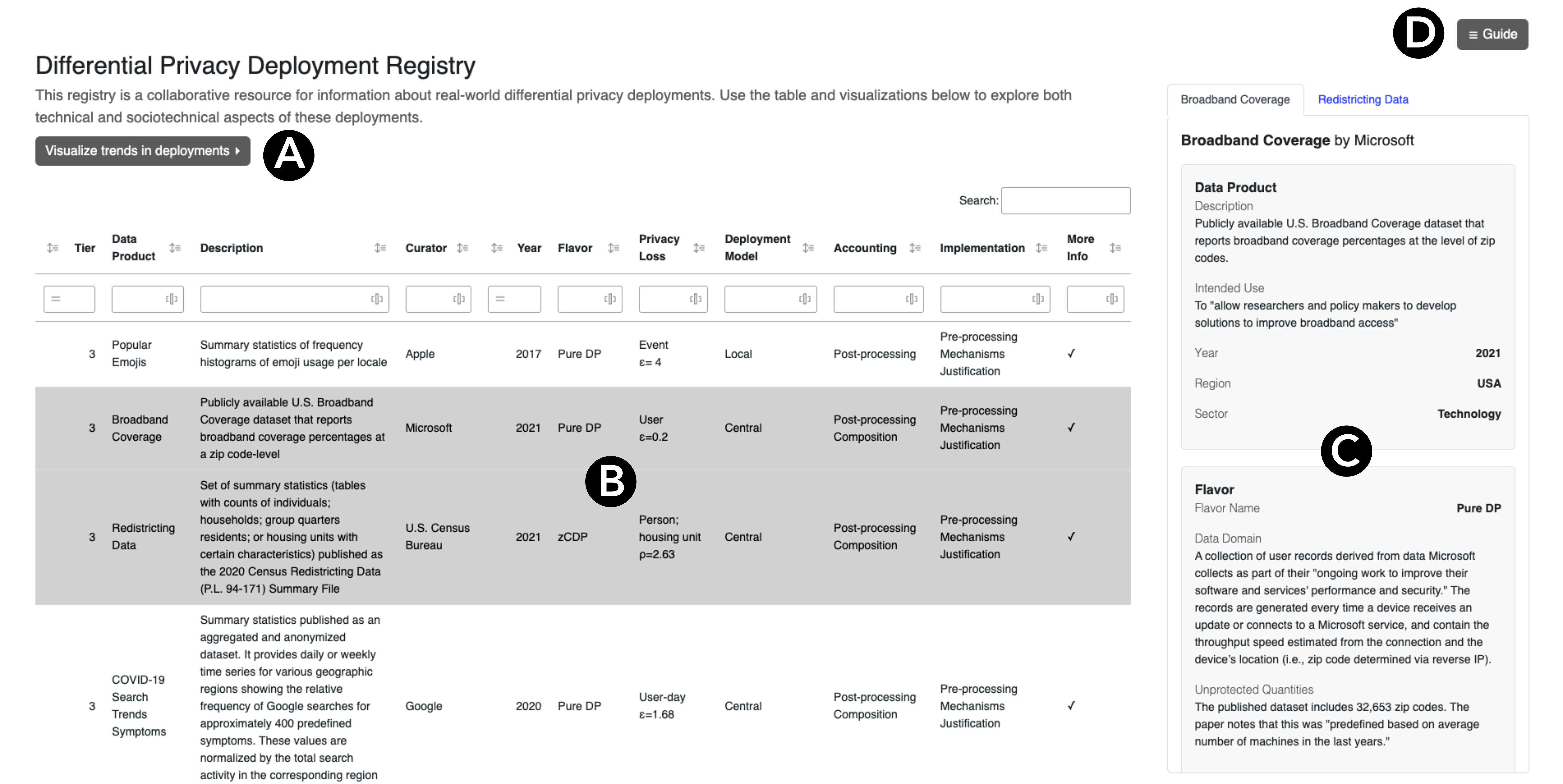}
    \caption{\footnotesize The registry as an interactive interface. The user has selected deployments by Microsoft~\cite{pereira2021us, MicrosoftUSBroadbandUsagePercentages} and the U.S. Census Bureau~\cite{abowd20222020, censusHandbook}.
    }
    \label{fig:registry}
\end{figure}

\begin{figure}[]
    \centering
    \includegraphics[width=.75\textwidth]{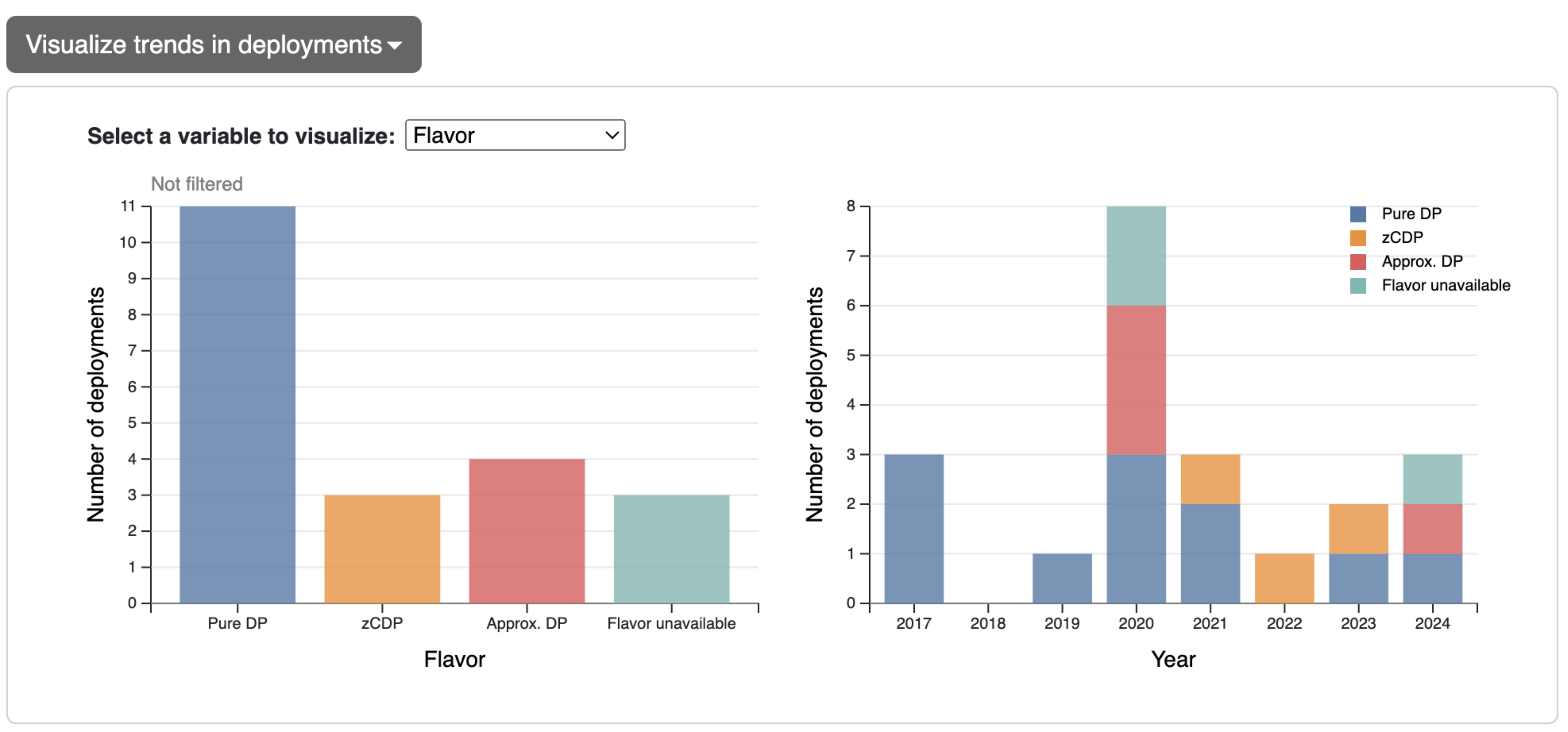}
    \caption{\footnotesize Exploratory, interactive visualizations that appear when clicking the ``Visualize trends in deployments'' button. The user has selected ``Flavor'' as the variable to visualize.
    }
    \label{fig:vis}
\end{figure}

\subsection{Interface components}
Below, we describe components of the interface.

\textbf{Table of deployments.} The interactive table (Figure~\ref{fig:registry}\textbf{B}) shows high-level information about deployments (\textbf{DG2}); each row represents a single deployment. The columns in the registry roughly correspond to sections in the deployment card described in Section~\ref{section:deployment_card}, with added columns for tier, description of the data product, and curator. The columns ``Implementation'' and ``Accounting'' contain keywords denoting whether more information on those topics is available in the deployment's card, described below. The column ``More info'' contains a check mark (\checkmark) if there is more information or sources available in the card. To help users identify and compare across similar deployments (\textbf{DG3}), columns are sortable, searchable, and filterable. The table also includes a search bar over the entire table so users can search a particular term without knowing which column in which it is likely to appear.\looseness=-1

We populated the registry with 21 real-world deployments: ten at Tier 3, eight at Tier 2, and three at Tier 1. We aimed for a mix of deployments across type of curator (large companies, startups, government organizations, etc.) and type of deployment (summary statistics, machine learning model, etc.). Either the first or second author filled out each deployment according to the schema---using publicly-available materials, such as papers and blog posts---and flagged ambiguities along the way. The research team met to resolve ambiguities. Our goal was not to comprehensively document all DP deployments, but rather to curate a substantive enough selection for study participants to interact with. We were also limited by time and resources (e.g., how many deployments for which there are detailed enough papers or blog posts to fill out a Tier 3 entry).

\textbf{Deployment card.} To the right of the table is a panel for deployment cards (Figure~\ref{fig:registry}\textbf{C}). Selecting a deployment in the table displays its card. Up to three deployments can be selected at a time. The user can tab between and compare deployments (\textbf{DG3}). The deployment card has sections corresponding to the schema described in Section~\ref{section:deployment_card} (\textbf{DG1}, \textbf{DG2}); certain sections do not appear for Tier 1 and 2 entries.

\textbf{Visualizations.} Directly above the table is an expandable panel with two exploratory visualizations (Figure~\ref{fig:registry}\textbf{A}; Figure~\ref{fig:vis}). The user can select one of several variables (including but not limited to deployment model, region, sector, and flavor) to visualize. The bar graph on the left shows counts of deployments across each of the selected variable's levels. The stacked bar chart on the right shows counts broken down by the selected variable's levels, across years. The two plots are linked; brushing the x-axis (years) on the stacked bar chart filters the data shown on the first bar graph. Together, these visualizations allow a user to systematically explore trends in the registry~(\textbf{DG4}).

\textbf{Guide.} The interface also provides a guide describing main concepts in the registry. The guide appears as an overlay on the screen when the ``Guide'' button (Figure~\ref{fig:registry}\textbf{D}) is clicked. The guide contains a section corresponding to each of the column headers in the table, with explanatory material and relevant citations pertaining to each (\textbf{DG1}).
\section{Methods}

To answer our RQs, we conducted a user study where practitioners interacted with the registry as a technology probe~\cite{hutchinson2003technology}, followed by a semi-structured interview about challenges and opportunities of a registry. Given that the registry is not a well-established artifact, our study was exploratory in nature. That is, our goal was to understand the space of possible perceptions surrounding a potential registry. As such, participants were given open-ended tasks which aligned with our exploratory RQs.

\subsection{Participants}
We recruited 16 participants through our professional networks. We sent our recruitment message to our contacts who apply DP in real-world settings and asked them to share it with their networks. Prospective participants filled out an interest form about their prior experience. We invited participants to our study if they had experience making or evaluating a real-world DP deployment. Upon completing the study, participants received a \$50 gift card. The study was exempted by our university's IRB.

\subsection{Protocol}
The first author led study sessions, each lasting approximately one hour. We began sessions with a brief semi-structured interview to learn about participants' roles and prior experience with DP. Next, participants completed four open-ended tasks (described below) using the registry. We used a think-aloud protocol~\cite{wright1991use}, and asked follow-up questions along the way. We ended each session by conducting a semi-structured interview about the participant's experience using the registry as well as challenges and opportunities around the registry gaining adoption. Interview questions are in the appendix.

\subsection{Tasks}
We designed four tasks to help answer our RQs. Participants completed tasks in the order shown below.

\textbf{Task 1: Examining a known deployment.}
We asked participants to familiarize themselves with the registry and its deployments. Then, we asked them to choose one deployment they were familiar with prior to interacting with the registry and read about it. Next, we asked them to \textit{share whether and how their impressions changed, including whether they learned any new technical or sociotechnical details}. Finally, we asked them whether and how their understanding of DP or how it is deployed changed after interacting with the registry.

\textbf{Task 2: Evaluating deployments.}
We asked participants to focus on deployments they were not already familiar with. The, we asked them to \textit{choose one deployment that stood out as ``good''} (and explain why) and \textit{one deployment that stood out as ``poor''} (and explain why).

\textbf{Task 3: Making a new deployment.}
For this task, we asked participants to either consider a hypothetical DP use case (details of which are in the appendix) or a use case they were currently considering in their own work. We asked them to use the registry to \textit{determine whether DP is appropriate for the use case and, if so, how they would implement it}.

\textbf{Task 4: Exploring emergent norms.}
Finally, we asked participants to use the exploratory visualizations or table to \textit{reflect on norms or standards appearing to emerge in the DP ecosystem}. We prompted them for their reaction to the standards they identified and asked about how the registry would help or hinder the emergence of such standards.

\subsection{Analysis}
We followed a thematic analysis approach~\cite{braun2006using} to qualitatively analyze participants' sessions. The first author began by open coding recordings of each participant's session. Next, they iteratively grouped quotations and observations across participants into themes within each RQ. They allowed new themes to emerge throughout this process and followed the method of constant comparison to continually assess the stability of each theme. Then, the entire research team met to iterate on and refine these themes.
\section{Results}

We begin by describing participants' backgrounds, then present results for each of our RQs.

\subsection{Participants' backgrounds}
Participants in our study had a mix of experiences deploying and evaluating DP deployments in real-world settings. They held positions across industry and government organizations, in roles including technical lead, privacy engineer, research scientist, and academic researcher. They were based globally: four in Europe, one in Asia, and 11 in North America. Fourteen participants held a PhD, one held a master's degree, and one held a bachelor's degree. They were largely experts in DP, reporting an average familiarity with DP of 4.5 (median $=$ 5) on a 5-point semantic scale.

\subsection{RQ1: How the registry changes understandings of DP}
\textbf{Did not always change high-level understandings of DP, but consistently supported discovering low-level details.} For some participants, interacting with the registry changed their high-level understanding of DP and how it is deployed (P2, P5, P6, P11, P12). Specifically, the registry showed them that DP is applied more widely than they previously knew. For example, the registry challenged P2 and P12's belief that DP was only applied by the U.S. Census Bureau and large tech companies. P11, who said they had a ``passing understanding of DP'' likened the space of DP deployments to an iceberg, where previously they knew of only ``the tip of the iceberg'' but the registry showed them a ``whole other world of these different approaches.'' On the other hand, some participants' (P1, P3, P7, P9, P14, P16) high-level understanding did not change after interacting with the registry, which is perhaps unsurprising given that many participants were already familiar with the deployments in the registry. However, even when participants said their overall understanding did not change, they reiterated how helpful the registry is for reminding them of details and having everything in one place instead of searching across papers, blog posts, and GitHub repositories: ``But this is so good as a resource...I'm happy. I'm smiling...It's very helpful to have a place where it's just easy to find...the important information'' (P9).

Participants consistently learned new details about individual deployments that they were already familiar with prior to interacting with the registry (P2, P4, P5, P8, P10--13, P15, P16), particularly after inspecting deployment cards. These details spanned both technical choices---such as details about unprotected quantities, hyperparameter tuning, algorithms, and DP flavors---and sociotechnical information, like rationale for deploying DP or the intended use of a data product. For example, P4 and P11 learned about the specific algorithm used for Apple's use of DP to protect histograms of emoji usage~\cite{apple2017} (Tier 3). As P13 explored the same deployment's card, they described learning about technical details as well as the purpose of the release:

\begin{quote} I wasn't clear on what their budgeting looked like. So now that has been cleared up. And then it's interesting to see...some of the other security measures for protecting the data. Okay, so it's to improve predictive emoji suggestions. So that I didn't know...Everyone just kind of talks vaguely about Apple using DP to protect emojis. So that's interesting to see. \end{quote}

P10's learning also focused on sociotechnical aspects. When reading the ``Justification'' subsection of the deployment card for the U.S. Census Bureau's redistricting data files~\cite{abowd20222020, censusHandbook} (Tier 3), they said they had not previously known that the bureau ``ran [an] attack on the previous release and it was this successful,'' referring to some of the re-identification rates that prompted the bureau to apply DP for the 2020 Census~\cite{censusHandbook}.

A few participants (P2, P5, P16) even described gaining new knowledge from Tier 2 deployments. These participants reported learning about either the region of the deployment or privacy parameters for a couple Tier 2 deployments with which they were already familiar. While participants understandably gained more knowledge from Tier 3 deployments, it seems that practitioners may even learn from reading about deployments that do not meet such a high transparency level, suggesting the usefulness of Tier 2 and possibly Tier 1 deployments.

\textbf{Inspired specific questions.} Participants found that reading the deployment cards made them curious about specific aspects of deployments and gave them leads to follow up on (P3--6, P8, P12, P15). These questions touched on a mix of technical and sociotechnical topics.

For example, when looking at one of Microsoft's deployments~\cite{MicrosoftUSBroadbandUsagePercentages, pereira2021us}, P15 wanted to know more about ``this ability to look at devices that are not using Microsoft services'' saying they were ``intrigued by how that's possible.'' P6 noticed that one of the Wikimedia Foundation's releases~\cite{adeleye2023publishing} did not collect persistent identifiers for users, saying that it would be useful to understand this process better, because ``[i]f your identifier is changing, potentially your privacy loss is increasing,'' and they ``don't think this is...a major problem, but potentially it could be if not done with a little bit of care.''

Participants also became curious about parts of deployments beyond specific implementation choices. For instance, P3 wanted more background on \emph{why} certain decisions were made:

\begin{quote}
    I feel like there's a lot of information here about concrete design choices. There is information about the privacy loss, etc., which is helpful and useful, but kind of the bigger or most interesting question is like, What led to that? What was...the risk they were worried about in a more concrete way? What...kind of...utility [did they want] to achieve?
\end{quote}

P5 noticed that the intended use of one deployment was to support public health authorities in decision-making during the COVID-19 pandemic, but had follow-up questions about which public health authorities in particular. P4 became curious whether a particular data product, which was used internally at an organization, was deleted after a certain period of time. Together, these examples suggest value in a registry that provides contextual information alongside specific implementation choices, echoing Dwork, Kohli, and Mulligan's~\cite{dwork2019differential} original call.

\subsection{RQ2: Using the registry}

\subsubsection{RQ2a: Evaluating deployments}
\textbf{Considered societal relevance plus various implementation choices.}
While common narratives around evaluating DP deployments emphasize privacy parameters, participants in our study demonstrated a more holistic approach to evaluating deployments. When asked to choose examples of ``good'' and ``poor'' deployments, participants factored in both overall societal relevance plus a range of fine-grained implementation choices.

Multiple participants (P3, P6, P7, P10--12) explicitly considered whether the deployment would help society and whether applying DP was necessary. As P3 put it, ``If you're using DP for...evil purposes, so what [if] you're protecting privacy, right?'' For them, evaluating a deployment entails an ``ethical assessment'' around whether it is worthwhile for society to have a particular data product available. P6, P10, and P12 all considered how well-suited DP was to the data. For P10, this meant paying particular attention to the ``Justification'' subsection of deployment cards to learn whether privacy protection was necessary in the first place. Their evaluation approach entailed first determining whether DP seemed necessary, and only then looking into technical details. Similarly, when identifying poor deployments, P12 looked for deployments where ``DP is an overkill'' noting that ``if an application does not necessarily require such stringent data privacy protection, that could be a bit of a flag for not a great deployment.'' P6 further elaborated that certain types of data, like browsing histories, were sensitive enough to warrant DP protections. They also considered whether DP as a technology was a good fit for the type of data release. For example, when evaluating one of the Wikimedia Foundation's deployments~\cite{adeleye2023publishing}, they felt that the data were a good fit for DP since the deployment was comprised of ``pageview counts per country...as opposed to highly rich crosstabs.''

Participants also considered various implementation choices (P1--5, P6, P7, P11--14, P16). Unsurprisingly, participants paid substantial attention to the privacy unit and privacy parameters, often by looking through the privacy loss column in the table. For example, when searching for a good deployment, P1 immediately scanned the table for small privacy parameters before looking more closely at other implementation decisions and trust assumptions. By and large, participants did not consider privacy parameters without the corresponding privacy unit. P12 explained the importance of the privacy unit: ``if we're looking at the privacy guarantee that is in this user-day style...that means that if an adversary can observe the...users' data over time...the privacy guarantee is going to eventually become vacuous.'' Given the importance they placed on the privacy unit and privacy parameters, participants (P3, P12, P14) wanted more direct ways of comparing between and interpreting these values in the registry. To this end, P12 and P14 said they wanted to see the full privacy profile, i.e., all ($\epsilon$, $\delta$) pairs. P12 explained that the full privacy profile---which cannot be deduced from an approximate DP guarantee which only provides one ($\epsilon$, $\delta$) pair---makes it easier to translate the guarantee into real-world risks:

\begin{quote}
    I [care] a lot about mapping DP to some sort of legible risk notions and approximate DP, where we only know one pair of ($\epsilon$, $\delta$) is a really poor representation of privacy guarantees when we map them to some notions of attack risks...That's why zCDP is better, because from there we can get a whole privacy profile.
\end{quote}

A few participants explicitly factored in utility as well (P1, P6, P15, P16). For example, P16 explained the difficulty of judging privacy-loss parameters, saying that ``...some privacy [parameter values] would just give garbage utility compared to others,'' suggesting that utility plays a role in their assessment of privacy-loss parameters and that an explicit section for utility may be beneficial to add to the schema. Because deployment cards in the registry largely do not contain explicit measures of utility, these participants sometimes made guesses about utility based on factors like the population size of the underlying data. For example, P1 stated that ``if [they] see something that has a very strong privacy guarantee and not a huge population, then [they're] concerned.''

In addition to the privacy unit and privacy parameters, participants considered factors like the deployment model, trust assumptions, and libraries used for implementation. For example, P14 looked not only at privacy loss, but also the deployment model. Specifically, they preferred deployments that limit the number of people with access to the raw data: ``...obviously if you can achieve the same $\epsilon$, but somehow you achieve it without having the server or many engineers...inspect the messages, that is great.'' They also looked at detailed implementation choices in the deployment cards, noting that they prefer when deployments use well-known DP libraries and open source their code. Along these same lines P5 and P14 both regarded a deployment more highly if it was possible to vet the details shared by the curator. As P5 looked through deployment cards, they specifically wanted to see ``Wikipedia style referencing where pretty much every claim has a pointer to where it came from,'' in addition to the ``More Info and Sources'' section at the end of each card. Such a referencing style may be beneficial to incorporate into a future version of the registry.

\textbf{More transparency was perceived as better.} Participants considered deployments with more available information to be better (P2, P4, P5, P9, P13, P11, P16). For some participants, transparency was a matter of principle. For example, P9 said that for them, the goodness of a deployment is ``related to the amount of details that are disclosed,'' further explaining that deployments that do not provide implementation details ``[feel] incomplete.'' P16 added that transparency is related to quality:

\begin{quote}
    ...one of the biggest guides for whether it's a good or bad deployment is how transparent they are. So the more information they publish, like code, data, and then discussions of limitations---...assuming technical correctness---that would be a good deployment. And so the less information that they provide, I guess the lower the quality is.
\end{quote}

When looking through a Tier 1 deployment, which contained only basic information about the data product and data curator, P5 commented ``the less information there is...the worse the deployment because the less information there is for even for an expert to judge what kind of risk [it poses].'' They later said that when no information was provided about a particular aspect of the deployment, they assumed a worse decision was made.

Without even clicking on deployment cards, P11 noticed deployments' level of transparency in the table itself, commenting that ``seeing these empty columns on the bottom [rows] makes me think maybe those are lower quality,'' suggesting a clear relationship between tiers and quality from participants' perspectives. When choosing a good deployment, P2 again focused on tiers: ``I'm kind of leaning towards tier 3. The fact that there's a lot of information around these also...gives me a good impression.''

We note that participants were explicitly briefed on the fact that tier assignment was not a measure of quality, but rather a measure of transparency. Therefore, the fact that participants drew a connection between tier and quality suggests transparency played an important role in how participants evaluated deployments.

\subsubsection{RQ2b: Informing future deployments}

\textbf{Did not rely on the registry for determining appropriateness of DP.} Before making lower-level implementation choices, practitioners must first decide whether DP is appropriate for a given use case. When given a hypothetical use case or considering one of their own (Task 3), participants tended not to rely on the registry for deciding whether DP would be appropriate (P3--5, P7, P11, P12)---despite being explicitly asking them to use the registry for this task. Instead, they relied on their prior knowledge about when DP is well-suited to a task, and considered how well attributes of the use case fit their understanding of when to apply DP. Sometimes, this meant that participants reflected on DP's appropriateness without returning to the registry. For example, when reading the hypothetical use case, P3 noticed that membership inference attacks~\cite{shokri2017membership}---which DP protects against---would be concerning, as knowing someone visited the website described in the use case could be a privacy violation on its own. Based on this observation, they decided DP should be applied for the data release. P4, P7, P11 all noticed that the size of the hypothetical dataset was large enough that DP would be effective (in terms of preserving utility), and similarly made their decision without consulting the registry. 

Some participants even explicitly stated that they would not use the registry for determining appropriateness of DP. For instance, P7 said that they already ``know that DP is appropriate for this kind of use case...because it passes [their] litmus test'' for when to use DP. Similarly, P5 stated that they ``have enough knowledge to make this decision without the registry,'' but also said they could imagine people with less DP expertise using the registry for this task.

\textbf{Used ``similar'' deployments as reference for implementation choices.} When asked to talk through how they might use the registry to make implementation choices for either the hypothetical use case described in Task 3 or a use case from their own work, most participants looked to the registry to first find similar deployments (P3, P4, P6, P8, P10--13, P15, P16). However, their understandings of ``similarity'' varied. For a few participants (P3, P12, P13), similarity was based on the type of data product, like summary statistics or machine learning model. To this end, P3 searched the data product description column for ``summary statistics'' while P12 searched for ``microdata.'' For others, similarity was based more on the underlying data to be protected (P8, P10, P11, P13, P15). For instance, as P8 considered the hypothetical use case, they explained that as ``the first step, [they] would try to...find deployments in the same domain, which is either surveys with demographic information or relatedly, anything health related.'' Participants also focused on the size of the data product (P6, P8), and in P8's case, both the ``scale of the input and also the scale of the output.'' Both these participants wished they could filter deployments by dataset size, which the registry does not support. A future version of the registry may include size of the dataset as an additional attribute of each deployment, ideally measured by approximate number of privacy units.

After finding similar deployments, participants zeroed in on specific implementation choices, like the mechanism used, decisions about post-processing, DP flavor, privacy unit, and privacy parameters. P8 spoke about the value of a reference point for privacy parameters in particular:

\begin{quote}
    ...people have a hard time choosing privacy parameters...we can say, okay, this data is of similar sensitivity to what we're looking at, and if we're having a lower $\epsilon$ than they have, then that's a sign that we're doing something right.
\end{quote}

To a similar end, P14 described how they would attempt to choose privacy parameters that placed them in the ``top 5\% or top 10\% if possible, as long as...it's achievable from an accuracy perspective.''

Beyond making specific implementation choices based off other deployments, participants (P6, P13, P14) also said they would use the registry to find libraries or code. P6 and P13 thought the registry could be helpful in finding code that was directly reusable, while P14 felt that seeing other practitioners' choices of libraries could help them assess which libraries the community was gaining consensus around, and therefore which they should use.

\subsubsection{RQ2c: Reflecting on norms}
\textbf{Approached trends with curiosity, but healthy skepticism.} Participants were often eager to explore patterns in the registry, either by interacting with the visualizations or simply scanning the table. Often, this exploration took the form of studying the prevalence of features like DP flavor, deployment model, interactivity, release type, and sector of the deployment, then reflecting on whether the trend they identified aligned with their existing knowledge (P1-4, P7-11, P13, P15, P16). By inspecting the visualization for deployments by sector, multiple participants (P1, P4, P13, P15, P16) noticed that most deployments were in the technology sector, which they found unsurprising. In P15's words, ``it makes sense that the people who are going to deploy DP are probably going to be the people who understand DP and have access to data.'' Some participants (P2, P7, P9, P16) were surprised to see that pure DP was the most represented flavor, while P1 further noticed that other flavors like approximate DP and zero-concentrated DP were becoming more popular over time. They reasoned that data curators ``don't want to see a drop in utility, even if it comes with a very strong privacy guarantee.'' In this way, observing trends in the registry prompted reflection among participants. Participants engaged in similar reflection around the fact that central DP was used more often than local DP, particularly over time (P1, P9, P10, P15).

At the same time, participants took their findings with a grain of salt, noting the small and biased sample represented in the registry (P2, P4, P7, P8, P13):

\begin{quote}
    ...I'd be mindful of...extrapolating from 21 self-reported deployments and honestly, I wouldn't read too much into these numbers... extrapolating anything from a small sample---even if it's perfectly collected---is somewhat...irresponsible. And then I'd be very mindful of the fact that these are deployments that we know about, and that are promoted [by] their respective companies...
\end{quote}

The non-representativeness of the deployments became salient as participants noticed the relatively few machine learning deployments in the registry (P2, P3, P5, P7). P2, for example, reasoned that the lack of these deployments was a reporting issue, as they knew of large companies training models under DP. In particular, they thought companies may not report using DP for machine learning because they are not confident the protections are actually ``working'' in the same way they would be for aggregate statistics.

As participants interacted with the registry and reflected on norms, their curiosity extended beyond trends that were easy to study through the visualizations (P1-3, P13, P16), suggesting opportunities for the registry to be expanded. Both P1 and P13 wanted to drill down on trends only within deployments in the central model and deployments whose data products were summary statistics, respectively. As the registry grows, it may make sense for visualizations to include additional faceting, allowing users to study more detailed cross-tabs. When P13 noticed an increase in health-related deployments in 2020, they suggested it would be helpful to see some narrative interpretation around the reason for such trends (e.g., the COVID-19 pandemic). Taking a step back, P2 said the visualizations prompted them to think about how often an organization that says they are protecting privacy are using DP---a question they quickly realized the DP-focused registry does not support answering. These findings suggest that the visualizations were useful in not only providing information to participants, but also in learning about the kinds of ecosystem-level questions they would hope to answer when using a registry.

\textbf{Over time, the registry could help with the emergence of norms.} Participants felt that the registry would aid the emergence of standards for DP deployments by increasing transparency around information that was previously siloed or kept private (P1, P2, P6, P9, P14, P15). In fact, P1 and P6 alluded to how helpful previous efforts to collate prior deployments have been, and the ``extra credence and credibility'' (P1) the registry would provide. In particular, P1 said that within their organization, practitioners have created ``their own kind of ad hoc lists of examples...to build an internal case for deployment'' and also referred to other blogs with similar lists. P6 specifically mentioned how knowing about previous deployments facilitates comparison, helping with norm building:

\begin{quote}
    I definitely think [the registry] will help in the setting of these norms, again, because people will compare. They'll compare, you know, what did this company do? What did that company do? How does that compare to what the government is doing? ... I've definitely seen [Damien Desfontaines' list of real-world deployments] referred to quite a bit as a resource to compare to. So, I could imagine that this registry would similarly get a lot of attention.
\end{quote}

There is also a risk of practitioners basing deployment decisions off prior deployments, especially if these deployments made suboptimal choices. Both P9 and P16 pointed out that the registry could steer people into making suboptimal choices if those seem to be popular among prior deployments. As P16 explained:

\begin{quote}
    So [the registry] will summarize what's popular, but what's popular is not necessarily what's the most principled or best approach for an application...there's no interactive discussion of the norm. It's just observation and then trying to fit in... Like, why did we do this? Well, because everyone else was doing it.
\end{quote}

On the other hand, P2 suggested that the registry would give organizations ``visibility for doing the right thing,'' which may incentivize practitioners to make better decisions.

\subsection{RQ3: Opportunities and challenges}

\textbf{Opportunity: Inform future implementation choices.} Participants felt the registry could be a helpful resource when making future deployments (P1-4, P8, P12-14, P16), particularly by helping inform implementation choices. At a high level, P12 noted that it would be helpful to see similar deployments which could act as a ``blueprint for [their] use case.'' Some participants (P1, P2, P4, P8) thought the registry would be specifically useful for setting privacy parameters, or at least coming up with proposals for values to discuss with their teams. To this end, P4 felt it would be useful to see the distribution of privacy parameters conditioned on the privacy unit. 

While the registry currently plays a predominantly descriptive role, participants suggested it should also adopt a normative role to support future deployments (P3, P4, P12). That is, they felt that it would be beneficial to flag libraries with known issues or provide some authoritative guidance around how one \textit{should} make certain implementation choices. For example, P4 suggested that along with listing privacy parameters, it would be helpful for the registry to include a note about common privacy parameters or ranges of ``very strong'' values, as determined by an entity like NIST. They went on to say that this kind of information ``could help inform other people's decisions.'' That said, consensus around parameter values largely does not exist, and would need to be established before being incorporated into the registry.

\textbf{Opportunity: Support broader communication about DP.}
Participants saw multiple opportunities for using the registry for broader communication about DP. Most notably, they thought the registry could help them advocate within their organizations for the use of DP (P1, P6, P7, P9, P15), particularly to policy or legal teams. For example, P9 stressed that being able to show internal legal teams other examples of deployments would help convince them that DP should be used, given the importance of precedence in legal contexts. P6 and P15 spoke about using the registry in conversations with teammates in managerial roles who presumably do not have deep DP expertise but hold decision-making power. In this way, it seems that participants envision the registry as having potential as a boundary object~\cite{star1989structure, star1989institutional} that can facilitate conversations between people in their organizations with different backgrounds and understandings of DP.

Participants also saw opportunities for the registry to facilitate public engagement with DP more broadly (P3, P5, P12, P13, P15), again by those in policy roles, but also by members of the public more generally. For example, P3 said that they could see the registry being used by regulators who want to recommend the use of PETs and could specifically ``empower a regulator to think...in a way that is more grounded with reality.'' Importantly, they noted that the registry in its current form may not be legible to a policy audience---presumably due to the level of technical jargon---but they could nevertheless envision the registry's potential for supporting this audience. P12 and P15 saw potential for the registry to help data subjects understand privacy protections. P15 thought the registry could be used to facilitate conversations between data curators and data subjects or others about how data should be protected. In other words, the registry could become a place to prospectively request feedback on deployment choices. 

Finally, participants saw potential for the registry to support various educational purposes (P5, P9, P11, P12). For example, P12 spoke about the educational role the registry can play in informing practitioners about how DP is deployed for a range of applications, while P5 mentioned it would ``be a fascinating tool for teaching'' and would help students ``get a holistic overview of what's going on and to then choose to dig deeper.''

Together, these findings indicate that although the registry in its current form is geared toward practitioners, there is substantial opportunity for it to support a range of other audiences in interacting with DP.

\textbf{Opportunity: Create standardization and community.}
Participants envisioned the registry as a place for the DP community to gather, interact, and foster shared language around DP deployments (P2, P5, P7, P11, P13-16). For example, P11 saw opportunity for the registry to become a ``hub'' for the community:

\begin{quote}
    ...the big [opportunity] would just be...having a hub with some standardized data. Because there's been these releases from...lots of different curators, lots of different sectors. I think it's really useful to just have it all in one place. And the opportunity...is just having...a standardized language on how DP releases work. And yeah, I guess establishing some norms.
\end{quote}

Similarly, P5 said the registry could be a ``go-to resource'' while P13 envisioned it as a ``go-to location for people...to find solutions with DP, but also to understand how DP is affecting the data that are used to drive social change that affects everybody.'' P7 and P15 even saw opportunity for practitioners to get feedback on their deployments or the way they report on their deployments. For instance, P7 imagined that a practitioner making their first DP deployment might have questions about how to report on their deployment, both in the registry and beyond, and might use the registry to find experts to guide them. Based on participants' comments, there seems to be opportunity for the registry to become a hub where members of the DP community can interact with one another, warranting further research into how such interaction should be facilitated.

\textbf{Challenge: Effort and risk involved with adding new entries.} In order for the registry to grow and become a reliable, up-to-date resource, it is critical that practitioners add new deployments to the registry. However, participants identified potential barriers to adding new entries, both in terms of the \textit{risk} incurred by making deployment details public and \textit{time and energy} required to add a new deployment.

First, participants felt that adding a deployment to the registry could bring about risks that may not outweigh possible benefits (P1, P6, P7, P9, P10, P14). P9 described the risk in terms of bringing extra scrutiny to a deployment, or opening the data curator up for regulatory punishment:

\begin{quote}
    So on one hand, I feel like transparency is key. On the other hand, I know that by sometimes...being too transparent, you can pay a price...people can come to the registry, understand how the data release was done and...criticize certain choices...But also, I know that from a regulation standpoint,...anything that...you're too transparent about can be eventually used against you.
\end{quote}

Similarly, P10 suggested that some organizations may not want to reveal details of their DP deployments because that would reveal that they are, in fact, collecting and using sensitive data, which the broader public may not be aware of otherwise. As P10 stated:

\begin{quote}
    ...sometimes in companies, they don't necessarily want to publish all use [cases] of privacy preserving technologies. Sometimes it's too sensitive...And yeah, by publishing it, you're kind of pointing out, oh, yeah, we are using that data, right? Even though we apply privacy preserving technology on top, right?...that doesn't sound very nice, right?
\end{quote}

To this end, P7 felt that within organizations, there would be ``internal stakeholders who...are going to see [adding a deployment to the registry] as a liability rather than a potential positive.'' Furthermore, P14 felt the registry should not encourage comparisons between or ranking of deployments, because then organizations would not add new entries:

\begin{quote}
    I think that some people may feel that the bar is high...if my implementation does not rise to the level that is set here, maybe I'm not going to include it because it's going to make me look bad...we don't want perfect to be the enemy of the good. And if we can somehow design this registry in a way that incentivizes people to ignore the fact that some deployments will be weak, that would be really good because it increases adoption.
\end{quote}

Second, participants thought that the logistics involved with adding a new entry could be a potential barrier (P7, P10, P11-13). For some, the time and energy required to add an entry did not feel worth it (P7: ``...people would be like, if I could snap my fingers and be on the registry with all the information, probably I would do it. But it would take me time and effort and I have better things to do.''). That said, P13 and P15 said they would add their own deployments to the registry, so whether effort required is a significant barrier in practice remains to be seen. For others, there were questions around how exactly to add an entry. As P8 commented, ``it's not clear to me how to add something to the registry. Do you write an e-mail, kind of fill out a form? What are the steps?'' In our study, we did not specify to participants a process for adding new entries, but these comments suggest that such a process should be clearly articulated, and made as simple as possible, to encourage adoption. 

\textbf{Challenge: Moderation.}
Participants raised several potential challenges related to moderation (P3, P5, P6, P9, P10, P12, P16). These concerns primarily centered around preventing false entries and keeping the registry updated. P5 summed up both these issues:

\begin{quote}
    ...who will have the ability to add these entries, right? And who will verify these entries?... How do you prevent [the registry] from becoming...spammed or inaccurate or out of date for that matter?
\end{quote}

P10 elaborated on the issue of keeping entries updated as parameter choices potentially change over time, for what might be considered the same deployment, and deployments may become inactive at any point (e.g., if they make many releases):

\begin{quote}
    I think, okay, in government [agencies], maybe it's not the case, but in companies, things move very quickly... some [projects] get started and get closed all the time. So it would be very hard to...keep track of this, right? ... But when you talk about, what is the [DP flavor], and what's the $\epsilon$, what's the level of DP, and so on, these things change very often, right? So who is going to come and report it...
\end{quote}

At the same time, participants had several ideas for how to facilitate moderation (P1, P2, P4, P8, P10, P11, P13, P14). In particular, multiple participants had thoughts on how new entries to the registry could be vetted for quality (P2, P5, P8, P10, P13, P16). P2, P5, P10, and P13 suggested various forms of expert review or maintenance over entries. For instance, P2 described a ``model like Wikipedia,'' where ``trusted editors...receive suggestions and go through details and validate'' them. P8 similarly referred to Wikipedia as a potential model for the registry, saying there should be ``someone ...with the maintainer status who keeps an eye on what's going on.'' To ensure long-term sustainability of expert review, P5 suggested raising money from large tech companies who deploy DP and using the funds to pay experts ``to vet the deployments based on the publicly available information.'' P5 went on to clarify that vetting information would entail the expert going through available documentation about the deployment to ensure that it is well-described in the registry.

There were also concerns about power dynamics regarding maintenance of the registry. P11 noted that they would suggest a ``communal, grassroots'' form of maintenance (``like a Wikipedia''), however they were also aware of challenges that might arise if the registry is dominated by ``huge companies and governments'' who may inadvertently end up ``governing the registry.'' To alleviate this issue, P14 suggested that the registry should be hosted and maintained by a ``nonprofit'' without ``incentives to make someone look good'':

\begin{quote}
    I think that team should be a nonprofit team, should not have incentives to make someone look good versus someone else. I would feel much better if it was academia, it was governed by professors and academics because I would feel that academia has no incentive to root for one deployment versus another. They kind of want the whole community to benefit. 
\end{quote}

Finally, participants offered ideas for making the registry more interactive. P4, for example, suggested allowing people to ``like or dislike'' deployments as a way of generating a ``community opinion'' about deployments. P1 and P14 spoke about the importance of allowing entries to be updated. While P14 thought corrections should be made publicly, through a ``method similar to GitHub,'' P1 felt that corrections would ideally be done through a ``back channel'':

\begin{quote}
    I mean, you could always make this more of a wiki or a bulletin board and have comment threads and so on. But I think that, if anything,... probably detracts from the sort of...authority of tone that [the registry] currently enjoys. So I think of it more as just that there's a contact point on each [deployment] that says, corrections or other concerns, e-mail here.
\end{quote}
\section{Impact and collaboration}
While conducting this research, we worked in collaboration with a larger team at OpenDP, Oblivious, and NIST to create a community-driven deployable DP registry~\cite{NIST.IR.8588.2025}. The work described herein informed the development of the registry intended for deployment; at the same time, there were aspects of mutual influence between that registry and the design described here.

When starting to develop our interface prototype, we used aspects of the prototype created by Oblivious~\cite{original_registry} as a starting point. As discussed in the introduction, in spring 2024 a team at Oblivious started to work toward a more systematic, community-driven registry to be further developed in collaboration with OpenDP. They created an initial prototype in blog post format with a table of deployments and educational content about DP. We expanded their basic set of attributes to create a holistic, hierarchical schema. This schema was then incorporated back into their registry, with a few minor modifications. There was mutual influence during the design of the interactive table and deployment card. For example, we shared early versions of an updated prototype and had shared discussions about further iterations, especially with regard to designing the deployment card. Finally, while the content in our ``Guide'' is largely distinct from the educational content in the initial prototype created by Oblivious, we used their text as a starting point.

Throughout our research process, we shared our interface, code for the visualizations, our schema, our Tier 2 and Tier 3 entries, and text from our guide with the larger team. Much of our design contributions and filled-out entries have been incorporated back into the registry intended for deployment. This registry was launched in September 2025 with a proposal by NIST to host it~\cite{NIST.IR.8588.2025}.
\section{Discussion and conclusion}
\subsection{Toward increased adoption of the registry}
To maximize the registry's long-term impact, practitioners must add new deployments as they are made. However, our results suggest that a big challenge to the registry gaining adoption will be the time and effort required for practitioners to add new entries, especially at the Tier 3 level. Thus, we propose strategies to make it easier to add entries.

First, it may be fruitful to develop partially automated, human-in-the-loop approaches for adding new entries. In this work, the first two authors manually filled out each deployment's entry. We also did some preliminary explorations into creating a large language model (LLM) agent for this task. Such an agent would extract information for the deployment card (from publicly-available documentation, like papers and blog posts), attach supporting quotes for each attribute, and indicate whether the information was explicitly stated, inferred from context, or unspecified, along with page, section, or figure references so that each value can be traced to its source. A practitioner, either involved in the deployment or with sufficient DP expertise, would then review and correct the suggestions before submitting to the registry. Review is essential, as legitimacy and credibility of the registry hinges on information being correct. Care should be taken to not only verify the agent's suggestions, but also to improve the agent based on practitioner feedback, for example, by adopting a dual-agent strategy where one model extracts information comprehensively (prioritizing recall) and another extracts information only when highly confident (prioritizing precision), so that overlaps can be accepted with minimal effort while disagreements are flagged for review. Ideally, registry maintainers would recommend a single LLM agent, and solicit feedback from practitioners after use so that it can be iteratively improved.

Second, it would be valuable for DP libraries and software to automatically save implementation details in a format that can be easily transferred to a deployment card. For example, once a differentially-private program is executed, the practitioner should then be able to download a file with attributes about the deployment (e.g., flavor, definition of adjacent datasets, privacy-loss parameters, composition, mechanisms). Some sociotechnical information, such as the intended use of the data product or rationale for implementation choices, will need to be manually filled out by the practitioner. Extracting information automatically from the implementation code has the added benefit of being able to verify correctness of the claims, which is impossible to do with only a paper describing implementation details.

Third, based on a participant's comment, we suggest creating incentives to add new deployments as part of conference publication procedures. For example, if a paper's contribution is a real-world DP deployment, authors may submit an entry to the registry as part of an optional artifact review, further lending credibility and exposure to their work. This would help incentivize adding entries among academic researchers, and would need to be translated to other types of venues to reach industry practitioners.

\subsection{Fostering community through a discussion board.}
Participants saw opportunity for the registry becoming a ``hub'' for the DP community, implying that the registry could become a meeting ground for practitioners. Currently, the registry is designed to be interactive insofar as practitioners can add new deployments to the registry, and users of the registry can interactively explore and compare deployments. However, there is significant potential to facilitate richer forms of interaction between community members. One basic change would be for each deployment to include contact information for someone who has agreed to answer questions about the deployment. This simple addition would also help foster a culture of question-asking, even between people who are not already acquainted. Several participants described how, when working on deployments, they often relied on either research papers or consultations (either formally or informally) with experts; for practitioners newer to DP, contact information would be especially helpful.

To enable further discussion, we propose adding to the registry a discussion board, which would allow community members to have back-and-forth discussions about deployments. These discussions would help standards emerge more intentionally. Participants in our study were concerned that precedents will drive future decisions, even if these precedents are not advisable. A forum would cultivate richer understandings of each deployment, including the extent to which its specific choices should be repeated. Thus, a discussion board would help create more normative guidance or interpretation, which participants in our study seemed to be seeking. Of course, a discussion board would raise a host of moderation challenges, which could be assessed during a trial period to assess benefits and costs.

\subsection{Making the registry usable beyond practitioners}
The original proposal for a registry~\cite{dwork2019differential} envisioned it as being useful not only for practitioners, but also for other parties, like policymakers interacting with DP. Our study further suggests that part of the registry's value is in facilitating communication with policymakers and data subjects. As is, our prototype is designed to support technically skilled practitioners.

As future work, we envision the registry being expanded to provide tailored support to policymakers, data subjects, and data users in interpreting implementation choices. In particular, the registry can be expanded to include several ``modes,'' each intended for a different party. For example, the mode for policymakers may provide interpretations of the deployments in their particular legal contexts and make connections between aspects of DP and various laws or policies~\cite{nissim2018privacy, cohen2020towards}. Support for data subjects may mean providing lay explanations of the privacy and accuracy guarantees offered by a particular deployment (see~\cite{dibia2024sok} for several approaches proposed by prior work). The extent to which these explanations can be automatically generated based on information already in the deployment card requires further research. Finally, when a data product is made publicly available, it would be helpful to support data users in understanding the implications of added DP noise on their downstream applications. Data curators may submit links to deployment documentation for data users alongside their registry entry. Such documentation may provide guidance on accounting for DP noise in downstream applications~\cite{triedman2025confidence}.
Together, these possible expansions can increase the registry's impact beyond practitioners.

\section*{Acknowledgments}
We thank our collaborators in OpenDP and Oblivious on the registry intended for deployment, including but not limited to Angela Barragan, Jack Fitzsimons, James Honaker, Chuck McCallum, Michael Shoemate, and Vikrant Singhal. In particular, we thank James Honaker for suggesting the idea of transparency tiers and Angela Barragan for helpful design discussions. We also thank Jessica Hullman for early discussions about the registry and Mandi Cai, Hyeok Kim, and Lydia Lucchesi for interface feedback. This work was supported in part by a grant from the Sloan Foundation to OpenDP.

\bibliographystyle{unsrtnat}
\bibliography{references} 

\section*{Appendix}
\section*{Pre-task interview guide}

\begin{itemize}
    \item Please describe your current role.
    \begin{itemize}
        \item How long have you been in this role?
        \item What are your primary responsibilities in this role?
    \end{itemize}

    \item How have you interacted with DP either in your current role or prior roles?

    \item Have you been involved with making a DP deployment? If yes, then:
    \begin{itemize}
        \item What kind of data were you seeking to protect? What domain was it in?
        \item What were your goals in deploying DP?
        \item What was the published data product?
        \item What were some challenges you faced?
        \item What resources did you use to inform your choices, if any?
        \item How did you communicate your deployment with external parties?
    \end{itemize}

    \item Have you previously analyzed or evaluated a DP deployment? If yes, then:
    \begin{itemize}
        \item Describe a DP deployment that you’ve analyzed or evaluated in the past.
        \item What were your goals?
        \item Which aspects of the deployment did you focus on in your analysis or evaluation?
        \item What resources did you use to support your analysis or evaluation, if any?
    \end{itemize}
\end{itemize}

\section*{Post-task interview guide}

\begin{itemize}
    \item What did you find challenging about using the registry? What did you find easy?
    \item What do you see as opportunities the registry could open up?
    \item What do you see as challenges to community-wide adoption of the registry?
    \begin{itemize}
        \item How might a registry gain legitimacy among practitioners?
        \item Do you see any potential barriers to new entries being added to the registry? For example, if you were to make a DP deployment in the future, would you add it to the registry?
    \end{itemize}
    \item What are some ways you think the value of the registry can be maximized through governance? In other words, how should the registry be governed by the community? How should new deployments be validated, if at all? How should changes to the registry — such as new pieces of information required to add a deployment — be vetted?
    \item How do you envision using a registry like this in your work in the future, if at all?
    \item Any other thoughts?
\end{itemize}

\section*{Task 3 hypothetical use case}
Your organization maintains a website with articles about different medical conditions. Your team has conducted a survey to find out who’s using the website, how they’re using it, and for what purposes.

You have received several tens of thousands of responses. Your team plans to publish aggregate statistics about the collected data.

These published statistics will be used internally at your organization to help improve the website. They will also be made available to researchers interested in studying questions about how people in different demographic or socioeconomic groups access medical information.

Sample survey questions include, but are not limited to:

\begin{itemize}
    \item In the last year, how often did you visit this website?
    \begin{itemize}
        \item More than once a week
        \item Once a week
        \item Once or twice a week
        \item Once a month
        \item A few times
        \item Did not visit the website in the last year
    \end{itemize}
    \item Which purposes do you use the website for? (select all that apply)
    \begin{itemize}
        \item To find information about medical conditions I have been diagnosed with
        \item To find information about medical conditions I have not been diagnosed with
        \item To find information about medical conditions friends or family have been diagnosed with
        \item To improve my general understanding about various medical conditions
    \end{itemize}
    \item Select up to two of the following website sections or features which you use most frequently.
    \begin{itemize}
        \item Symptom checker
        \item Support group finder
        \item Diseases \& conditions database
        \item Drugs \& supplements database
        \item Tests \& procedures database
    \end{itemize}
    \item What is your gender? (select all the apply)
    \begin{itemize}
        \item Female
        \item Male
        \item Non-Binary
        \item Prefer to self describe: \noindent\rule{3cm}{0.4pt}
    \end{itemize}
    \item Please specify your race / ethnicity. (select all that apply)
    \begin{itemize}
        \item Hispanic or Latino
        \item Black or African American
        \item White
        \item American Indian or Alaska Native
        \item Asian, Native Hawaiian, or Pacific Islander
        \item Prefer to self describe: \noindent\rule{3cm}{0.4pt}
    \end{itemize}

    \item What is the highest level of school you have completed or the highest degree you have received?
    \begin{itemize}
        \item Less than a high school degree
        \item Some high school credit, no diploma or equivalent
        \item High school graduate (high school diploma or equivalent including GED)
        \item Some college but no degree
        \item Associate’s degree
        \item Bachelor’s degree
        \item Advanced degree (e.g., master’s, doctorate)
    \end{itemize}

    \item Which of the following includes your total HOUSEHOLD income for last year, before taxes (in USD)?
    \begin{itemize}
        \item Less than \$10,000
        \item \$10,000 to under \$50,000
        \item \$50,000 to under \$100,000
        \item \$100,000 to under \$200,000
        \item \$200,000 or more
    \end{itemize}
\end{itemize}

First, use the registry to determine whether DP seems appropriate for this use case.

If you believe DP is appropriate: how might you use the registry to inform decisions about how to implement DP in this use case? What aspects of other deployments would you consider? Which deployments would you focus on the most?

\textit{Aspects of the hypothetical website described above were based on WebMD and the Mayo Clinic's website.}

\end{document}